\documentclass[doublecol]{epl2} 
\usepackage{graphics, graphicx}
\usepackage[dvipsnames]{xcolor}
\usepackage{hyperref}
\usepackage{ulem}
\hypersetup{colorlinks, citecolor = black, filecolor = black, linkcolor = black, urlcolor = black}

\title{Convective heat transport in slender cells is close to that in wider cells at high Rayleigh and Prandtl numbers}
\shorttitle{Convective heat transport in slender cells is close to that in wider cells at high Rayleigh and Prandtl numbers} 

\author{Ambrish Pandey\inst{1} \and Katepalli R. Sreenivasan\inst{1,2}}
\shortauthor{A. Pandey and K. R. Sreenivasan}

\institute{                    
  \inst{1}Center for Space Science, New York University Abu Dhabi, Abu Dhabi 129188, United Arab Emirates\\
  \inst{2}Department of Physics, Courant Institute of Mathematical Sciences, Tandon School of Engineering, New York University, New York, NY 11201
}
\pacs{47.55.pb}{Thermal convection}
\pacs{47.55.P-}{Buoyancy-driven flows; convection}
\pacs{47.27.te}{Turbulent convective heat transfer}

\abstract{ 
Direct numerical simulations of turbulent convection at high Rayleigh numbers in large aspect ratio cells are challenging because of the prohibitive computational resources required. One can achieve high Rayleigh numbers at affordable costs for low aspect ratios, but the effect of small aspect ratio remains to be understood fully. In this work, we explore integral quantities in convection in a cell with the small aspect ratio of 0.1 by varying both the Rayleigh and Prandtl numbers systematically. We find that the heat transport in this flow is within 10\% of that in cells with large aspect ratios for high enough Rayleigh numbers and for Prandtl numbers larger than unity. For low Prandtl numbers, the increase of the heat transport is steeper for low aspect ratios, approaching that in large aspect ratios as the Prandtl number increases. Further, the global momentum transport, quantified by the Reynolds number, is reduced for all Prandtl numbers, presumably because of the larger volume of flow affected by the friction from sidewalls, compared to that in cells of larger aspect ratio. 
}

\begin{document}

\maketitle

\section{Introduction}

Convective flows in nature are highly complex and subject to many influences. Their controlled paradigm is the Rayleigh-B\'enard convection (RBC), where the fluid placed in a wide container is heated from below and cooled from above~\cite{Tritton:book1977, Ahlers:RMP2009, Chilla:EPJE2012, Verma:NJP2017, Verma:book2018, Schumacher:RMP2020}; all the thermal energy supplied at the bottom  plate is transmitted to the top plate entirely through the fluid. In RBC, the Rayleigh number ($Ra$) characterizes the strength of the driving buoyancy force compared to dissipative and diffusive forces. It is large in all the important natural flows \cite{Sreenivasan:1998}---on the order of $10^{20}$ and larger in solar convection and on the order of $10^{18}$ in the Earth's atmosphere. For this reason, researchers have attempted to achieve such high Rayleigh numbers in RBC in the laboratory~\cite{Castaing:JFM1989, Niemela:Nature2000, Niemela:JFM2003, Skrbek:JFM2015}, the highest achieved so far being $10^{17}$ in a cylindrical cell of aspect ratio $\Gamma = 0.5$, where the aspect ratio $\Gamma$ is the ratio of the diameter to the height of the convection cell. Direct numerical simulations (DNS) have also been utilized to study the detailed flow properties and morphology of RBC~\cite{Ahlers:RMP2009, Verma:NJP2017, Verma:book2018, {Stevens:JFM2010}}. The accessible values of $Ra$ in simulations too are limited by the available computing power. The computational effort needed to simulate a flow decreases with decreasing $\Gamma$, so Iyer et al. \cite{Iyer:PNAS2020} explored RBC in a slender cylindrical cell of $\Gamma = 0.1$ and reached up to $Ra=10^{15}$. To understand the effects of this confinement, it is necessary to explore in somewhat more detail the nature of the flow as $Ra$ is varied. Iyer et al. computed convection for a fluid with $Pr = 1$, where the Prandtl number $Pr$ is the ratio of kinematic viscosity $\nu$ and thermal diffusivity $\kappa$ of the working fluid. The Prandtl number in natural flows is often different from unity, sometimes greatly so \cite{Chilla:EPJE2012, Schumacher:RMP2020}. As many properties of convection differ between low-$Pr$ and high-$Pr$ flow, it is also important to explore the effects of flow confinement at different Prandtl numbers. In this letter, we explore RBC in a $\Gamma=0.1$ cylindrical cell and find that the global heat transport, which differs from that in cells with moderate and large $\Gamma$ at low Rayleigh and Prandtl numbers, agrees to within 10\% with that in high$-\Gamma$ cells when $Ra$ is large and $Pr = O(1)$ and higher.

In RBC, turbulent heat transport is quantified using the Nusselt number $Nu$ and momentum transport by the Reynolds number $Re$, both of which are functions of $Ra, Pr$, and $\Gamma$. In RBC with $\Gamma \geq 0.5$, $Nu \sim Ra^{\alpha} Pr^{\beta}$ has been observed with $\alpha \in [0.25-0.33]$ and $\beta \in [0.11-0.19]$~\cite{Castaing:JFM1989,  Cioni:JFM1997, Glazier:Nature1999, Verzicco:JFM1999, Niemela:Nature2000, Niemela:JFM2003, Verma:PRE2012, Stevens:JFM2013, Scheel:PRF2017, Pandey:Nature2018}, with $\alpha = 1/3$ appearing to hold for high Rayleigh numbers. The occasional claims in favor of slightly larger values of $\alpha$ \cite{He:PRL2012, Zhu:PRL2018} have been disputed~\cite{Skrbek:JFM2015, Doering:PRL2019}. In Iyer et al.~\cite{Iyer:PNAS2020}, the power law exponent $\alpha$ as well as the prefactor of $Nu-Ra$ relation was found to be consistent with these results, but it is obvious that the severe geometric constraint could have some influence. The effects of confinement were studied in a cylindrical cell in \cite{Niemela:JFM2006} and in rectangular domains in \cite{Wagner:POF2013, Huang:PRL2013, Chong:PRL2015, Chong:JFM2016}. Wagner et al.~\cite{Wagner:POF2013} found for $Pr = 0.786$ that $Nu$ and $Re$ decrease when the lateral aspect ratio is decreased from 1 to 0.1, owing to increasing friction from the walls as the confinement is increased. Further, the suppression in $Nu$ and $Re$ were observed to decrease with increasing $Ra$~\cite{Wagner:POF2013}. In Refs.\ \cite{Huang:PRL2013, Chong:PRL2015, Chong:JFM2016}, it was observed that the lateral confinement of a rectangular cell filled with water up to the aspect ratio $1/128$ causes first the suppression and then the enhancement of $Nu$ before the eventual suppression. The heat transport enhancement was attributed to condensation of the thermal plumes as the flow is severely confined~\cite{Huang:PRL2013}. The momentum transport was found to decrease with increasing confinement due to a larger volume of the flow affected directly by the sidewall friction ~\cite{Wagner:POF2013, Huang:PRL2013, Chong:PRL2015, Chong:JFM2016, Iyer:PNAS2020}. In Wagner et al.~\cite{Wagner:POF2013}, it was noted that, in contrast to that in $\Gamma = O(1)$ RBC, the vertical temperature profile exhibits a temperature gradient even in the bulk region of the cell. These results are consistent with \cite{Iyer:PNAS2020} and also with \cite{Zwirner:JFM2018} in a cylindrical cell of $\Gamma = 1/5$. 

In this letter, we explore the scaling of the integral quantities in RBC for $0.005 \leq Pr \leq 200$ and $10^8 \leq Ra \leq 3 \times 10^{10}$ for $\Gamma = 0.1$. Briefly, we find that the heat transport properties to be close to that for large $\Gamma$ when $Ra$ is large. However, the exponent $\beta$ in the slender cell for $Ra = 10^8$ and $10^9$ are larger than the range cited earlier for convection in $\Gamma \geq 0.5$ cells, and approaches $\beta \approx 0.19$ for $Ra \geq 10^{10}$. We also find that the mean temperature gradient in the bulk region decreases with increasing $Pr,$ and becomes small for $Pr > 1$. 

\section{Direct numerical simulations}

We perform DNS of RBC in a slender cylindrical cell of $\Gamma = 0.1$ for $Pr$ between 0.005 and 200 and $Ra$ between $10^8$ and $3 \times 10^{10}$. The incompressible ($\nabla \cdot {\bm u} = 0$) equations governing the dynamics of the convective flow are 
\begin{eqnarray}
\frac{\partial {\bm u}}{\partial t} + {\bm u} \cdot \nabla {\bm u} & = & -\nabla p + T \hat{z} + \sqrt{\frac{Pr}{Ra}} \, \nabla^2 {\bm u}, \label{eq:u} \\
\frac{\partial T}{\partial t} + {\bm u} \cdot \nabla T & = &\frac{1}{\sqrt{Pr Ra}}\, \nabla^2 T, \label{eq:T}
\end{eqnarray}
where ${\bm u} \, [=(u_x,u_y,u_z)], T, p$ are the velocity, temperature, and pressure fields, respectively. These equations are non-dimensionalized using the cylinder's height $H$, the free-fall velocity $u_f = \sqrt{\gamma g \Delta T H}$, and the temperature difference between the plates $\Delta T$ as the length, velocity, and temperature scales, respectively. Here $\gamma$ is the isobaric thermal expansion coefficient of the fluid and $g$ is the acceleration due to gravity. We apply the isothermal condition on the horizontal plates and adiabatic condition on the sidewall. All boundaries satisfy no-slip condition for the velocity field. The governing equations are integrated using a spectral element solver {\sc Nek5000}, where the domain is decomposed into a finite number of elements and the turbulence fields within each element are spectrally expanded using Lagrangian interpolation polynomials~\cite{Fischer:JCP1997, Scheel:NJP2013}. The simulations are started from the conduction state with random perturbations and the data analysis is performed only after initial transients have decayed. We ensure the adequacy of the spatial resolution using the criteria summarized in Scheel et al.~\cite{Scheel:NJP2013}. The important parameters of all the simulations are summarized in tables~\ref{tab.1} and \ref{tab.2}.

\begin{table}
\caption{Details of DNS for $Ra = 10^8$ and $10^9$: $N_e$ is the total number of spectral elements and $N$ is the order of the Lagrangian interpolation polynomials within each element. The error bars in $Nu$ and $Re$ represent the difference of the mean values computed over the first and second halves of the data sets.}
\label{tab.1}
\begin{center}
\begin{tabular}{lccccc}
\hline
$Ra$ & $Pr$ & $N_e$ & $N$ & $Nu$ & $Re$ \\
\hline
$10^8$ & 0.005 & 537600 & 3 & $1.30 \pm 0.02$ & $6756 \pm 18$ \\
$10^8$ & 0.01 & 537600 & 3 & $1.77 \pm 0.03$ & $5724 \pm 10$ \\
$10^8$ & 0.021 & 192000 & 3 & $2.82 \pm 0.05$ & $4498 \pm 4$ \\
$10^8$ & 0.035 & 192000 & 3 & $4.03 \pm 0.01$ & $3693 \pm 1$ \\
$10^8$ & 0.05 & 192000 & 3 & $5.05 \pm 0.23$ & $3124 \pm 13$ \\
$10^8$ & 0.075 & 192000 & 3 & $6.86 \pm 0.03$ & $2629 \pm 1$ \\
$10^8$ & 0.1 & 192000 & 3 & $8.25 \pm 0.13$ & $2277 \pm 3$ \\
$10^8$ & 0.2 & 192000 & 3 & $12.3 \pm 0.3$ & $1549 \pm 2$ \\
$10^8$ & 0.35 & 192000 & 3 & $14.0 \pm 0.7$ & $1012 \pm 4$ \\
$10^8$ & 0.5 & 192000 & 3 & $17.9 \pm 1.5$ & $832 \pm 5$ \\
$10^8$ & 0.7 & 192000 & 3 & $26.3 \pm 0.0$ & $746 \pm 0$ \\
$10^8$ & 1 & 192000 & 3 & $28.2 \pm 0.0$ & $581 \pm 0$ \\
$10^8$ & 2 & 192000 & 3 & $37.5 \pm 0.0$ & $369 \pm 0$ \\
$10^8$ & 4.38 & 192000 & 3 & $35.1 \pm 0.0$ & $164 \pm 0$ \\
$10^8$ & 7 & 192000 & 3 & $31.5 \pm 0.0$ & $96 \pm 0$ \\
$10^8$ & 20 & 192000 & 3 & $33.8 \pm 0.0$ & $35 \pm 0$ \\
$10^8$ & 100 & 192000 & 3 & $33.5 \pm 0.4$ & $7 \pm 0$ \\
$10^8$ & 200 & 192000 & 3 & $33.6 \pm 0.0$ & $3.5 \pm 0$ \\
$10^9$ & 0.021 & 537600 & 7 & $12.9 \pm 0.7$ & $19823 \pm 60$ \\
$10^9$ & 0.07 & 537600 & 5 & $20.8 \pm 2.4$ & $9785 \pm 77$ \\
$10^9$ & 0.1 & 192000 & 5 & $25.0 \pm 1.4$ & $8054 \pm 38$ \\
$10^9$ & 0.2 & 537600 & 3 & $31.7 \pm 1.4$ & $5215 \pm 13$ \\
$10^9$ & 0.35 & 537600 & 3 & $40.5 \pm 1.9$ & $3754 \pm 10$ \\
$10^9$ & 0.5 & 537600 & 3 & $46.5 \pm 0.9$ & $3016 \pm 4$ \\
$10^9$ & 0.7 & 537600 & 3 & $51.8 \pm 1.5$ & $2416 \pm 5$ \\
$10^9$ & 1 & 537600 & 3 & $58.4 \pm 2.5$ & $1923 \pm 7$ \\
$10^9$ & 2 & 537600 & 3 & $71.7 \pm 1.2$ & $1204 \pm 2$ \\
$10^9$ & 4.38 & 537600 & 3 & $76.7 \pm 4.0$ & $638 \pm 2$ \\
$10^9$ & 7 & 537600 & 3 & $77.2 \pm 1.6$ & $422 \pm 1$ \\
$10^9$ & 20 & 537600 & 3 & $75.3 \pm 0.1$ & $146 \pm 0$ \\
$10^9$ & 100 & 537600 & 3 & $71.6 \pm 0.1$ & $29 \pm 0$ \\
$10^9$ & 200 & 537600 & 3 & $70.3 \pm 1.4$ & $14 \pm 0$ \\
\hline
\end{tabular}
\end{center}
\end{table}

\begin{table}
\caption{Details of DNS for $Ra = 10^{10}$ and $3 \times 10^{10}$.}
\label{tab.2}
\begin{center}
\begin{tabular}{lccccc}
\hline
$Ra$ & $Pr$ & $N_e$ & $N$ & $Nu$ & $Re$ \\
\hline
$10^{10}$ & 0.1 & 192000 & 9 & $67.4 \pm 11.5$ & $25802 \pm 323$ \\
$10^{10}$ & 0.2 & 192000 & 7 & $79.5 \pm 1.2$ & $16735 \pm 8$ \\
$10^{10}$ & 0.35 & 192000 & 5 & $88.4 \pm 1.5$ & $11466 \pm 15$ \\
$10^{10}$ & 0.5 & 192000 & 5 & $95.2 \pm 0.6$ & $9030 \pm 4$ \\
$10^{10}$ & 0.7 & 192000 & 5 & $99.8 \pm 1.8$ & $7118 \pm 10$ \\
$10^{10}$ & 1 & 192000 & 5 & $106.6 \pm 2.1$ & $5564 \pm 6$ \\
$10^{10}$ & 2 & 192000 & 5 & $120.9 \pm 1.4$ & $3435 \pm 1$ \\
$10^{10}$ & 4.38 & 192000 & 5 & $131.0 \pm 6.9$ & $1921 \pm 3$ \\
$10^{10}$ & 7 & 192000 & 5 & $135.0 \pm 1.3$ & $1332 \pm 1$ \\
$10^{10}$ & 20 & 192000 & 5 & $134.1 \pm 0.4$ & $538 \pm 1$ \\
$10^{10}$ & 100 & 192000 & 5 & $120.5 \pm 3.2$ & $98 \pm 1$ \\
$10^{10}$ & 200 & 192000 & 5 & $123.2 \pm 1.5$ & $50 \pm 1$ \\
$3 \cdot 10^{10}$ & 0.1 & 537600 & 11 & $104.2 \pm 8.3$ & $44362 \pm 382$ \\
$3 \cdot 10^{10}$ & 0.2 & 192000 & 11 & $111.0 \pm 17.9$ & $27653 \pm 228$ \\
$3 \cdot 10^{10}$ & 0.35 & 192000 & 9 & $131.9 \pm 6.6$ & $19748 \pm 43$ \\
$3 \cdot 10^{10}$ & 0.5 & 192000 & 7 & $134.3 \pm 9.2$ & $15173 \pm 64$ \\
$3 \cdot 10^{10}$ & 0.7 & 192000 & 7 & $140.9 \pm 17.2$ & $12043 \pm 81$ \\
$3 \cdot 10^{10}$ & 1 & 192000 & 7 & $152.8 \pm 1.2$ & $9492 \pm 7$ \\
$3 \cdot 10^{10}$ & 2 & 192000 & 7 & $160.6 \pm 5.9$ & $5650 \pm 9$ \\
$3 \cdot 10^{10}$ & 4.38 & 192000 & 7 & $178.1 \pm 3.7$ & $3237 \pm 5$ \\
$3 \cdot 10^{10}$ & 7 & 192000 & 7 & $181.7 \pm 14.3$ & $2287 \pm 8$ \\
$3 \cdot 10^{10}$ & 20 & 192000 & 7 & $182.0 \pm 7.1$ & $964 \pm 2$ \\
$3 \cdot 10^{10}$ & 100 & 192000 & 7 & $179.2 \pm 12.6$ & $207 \pm 1$ \\
$3 \cdot 10^{10}$ & 200 & 192000 & 7 & $179.4 \pm 2.6$ & $103 \pm 1$ \\
\hline
\end{tabular}
\end{center}
\end{table}

\section{Temperature profiles}

\begin{figure}
\centerline{\includegraphics[width=0.5\textwidth]{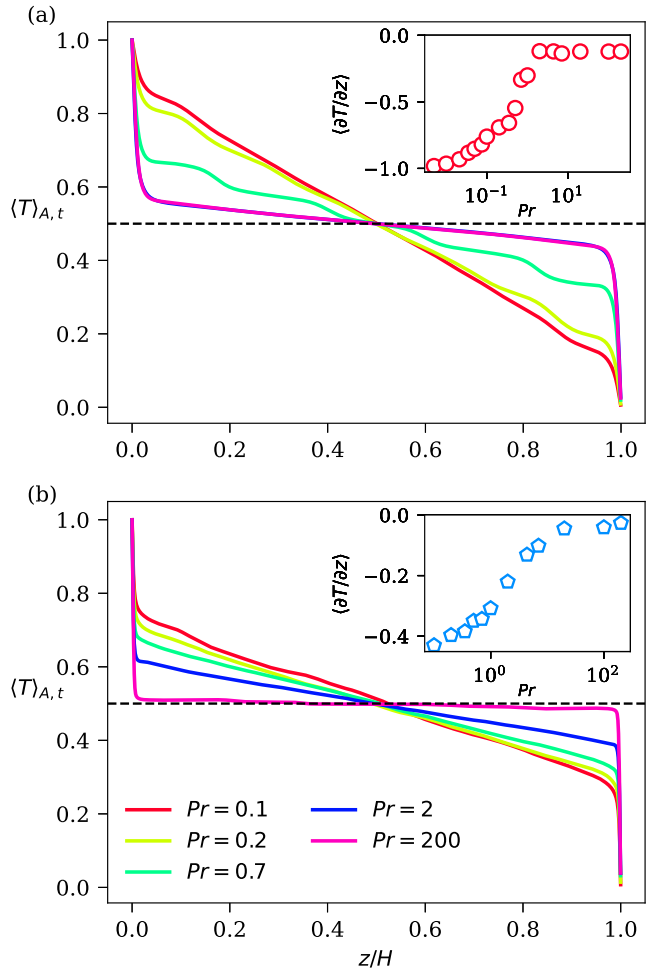}}
\caption{Horizontally- and temporally-averaged temperature as a function of the vertical distance from the plate for $Ra = 10^8$ (a) and $Ra = 10^{10}$ (b). The profiles exhibit a nonzero mean temperature gradient in the bulk, which decreases with increasing $Pr$ (as indicated in the insets, where the averaged $\partial T/\partial z$ in the bulk region is plotted as a function of $Pr$). The horizontal dashed line in both panels indicates the mean temperature 0.5 of the fluid. The profiles for $Pr = 0.2$ and 0.7 in panel (a) exhibit a wavy structure, which is due to multiple vertically-stacked rolls that persist for a long time due to the low Reynolds number. The corresponding vertical and horizontal velocity profiles also exhibit the wavy structure.}
\label{fig:T_z}
\end{figure}

In contrast to a large-scale circulation observed in $\Gamma \approx 1$ cells, we observe multiple vertically stacked convection rolls, whose number depends on $Pr$. Further, we do not observe very well mixed isothermal bulk region~\cite{Iyer:PNAS2020} in our cell for low and moderate Prandtl numbers. This is exhibited in fig.~\ref{fig:T_z}, where the horizontally- and temporally-averaged temperature profiles $\langle T \rangle_{A,t}(z)$ for a few simulations at $Ra = 10^8$ and $10^{10}$ are plotted. 
We note that $\langle T \rangle_{A,t}(z)$ for $Pr = 0.1$ and $Ra = 10^8$ (red curve in fig.~\ref{fig:T_z}(a)) varies almost linearly (outside the sharp drop near the plates) for the entire height of the cell. With increasing $Pr$, the temperature drop in the bulk region becomes smaller as exhibited in fig.~\ref{fig:T_z}(a). The profiles in fig.~\ref{fig:T_z}(b) indicate an increasingly mixed bulk region for flows at $Ra = 10^{10}$. We observe that the bulk temperature gradient remains significant for $Pr \leq 1$ even for the highest $Ra$ explored in the present work~\cite{Iyer:PNAS2020}. 

To quantify the variation of bulk temperature gradient with varying $Pr$, we compute the averaged vertical temperature gradient in the bulk region $\langle \partial T/\partial z \rangle$ by averaging over the entire horizontal cross-section and height between $z = 0.25H$ and $0.75H$. The insets of fig.~\ref{fig:T_z} show $\langle \partial T/\partial z \rangle$ as a function of $Pr$, which reveals that $\langle \partial T/\partial z \rangle \approx -1$ for $Ra=10^8$ and $Pr=0.005$, very close to the conduction value. Thus the conduction temperature profile is perturbed only slightly despite a highly turbulent bulk flow~\cite{Schumacher:PNAS2015,Schumacher:RMP2020}. For $Pr \geq 100$ and $Ra \geq 10^{10}$, however, we find that almost the entire temperature drop occurs within the thermal boundary layers (BLs) and the bulk flow remains nearly isothermal as indicated in the inset of fig.~\ref{fig:T_z}(b). 

\section{Heat transport}

\begin{figure}
\centerline{\includegraphics[width=0.47\textwidth]{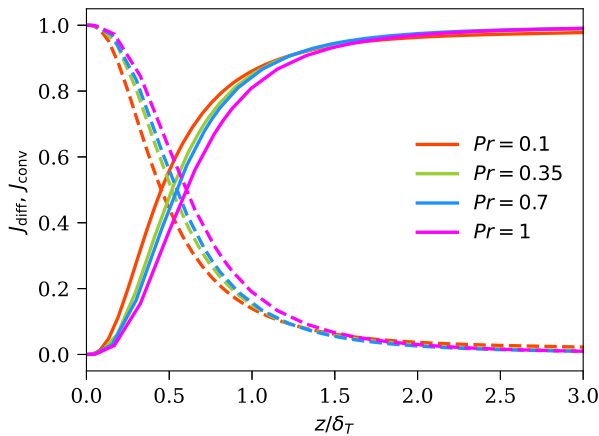}}
\caption{Convective (solid curves) and diffusive (dashed curves) fractions of the total heat flux near the plate for $Ra = 3 \times 10^{10}$ for various Prandtl numbers.}
\label{fig:Nu_z}
\end{figure}

In RBC, the total heat transport computed in a horizontal plane as
\begin{equation}
\label{eq:Nu_z}
Nu(z) = \sqrt{Ra Pr} \, \langle u_z T \rangle_{A,t} - \partial \langle T \rangle_{A,t}/\partial z
\end{equation}
remains independent of $z$. It is dominated by the convective fraction $J_\mathrm{conv} = \sqrt{RaPr} \, \langle u_z T \rangle_{A,t}/Nu$ in the bulk, and by the diffusive fraction $J_\mathrm{diff} = (-\partial \langle T \rangle_{A,t}/\partial z)/Nu$ within the thermal BLs. We show $J_\mathrm{conv}$ and $J_\mathrm{diff}$ as a function of the normalized distance $z/\delta_T$ from the bottom plate  in fig.~\ref{fig:Nu_z} for a few Prandtl numbers at $Ra = 3 \times 10^{10}$. Here, the thermal boundary layer (BL) thickness $\delta_T$ is estimated (for convenience) using the relation $\delta_T = 0.5H/Nu$, where $Nu$ is the globally- and temporally-averaged heat flux. We have verified that $\delta_T$ computed using the slope method~\cite{Pandey:JFM2021} agrees well with that estimated using the above relation.  Figure~\ref{fig:Nu_z} shows that the total heat flux at the plate is entirely due to the molecular diffusion. However, further away from the plate, the convective component increases and starts dominating the total heat transfer even within the BL region. Fig.~\ref{fig:Nu_z} further reveals that, for a given relative distance from the plate, $J_\mathrm{conv}$ increases and $J_\mathrm{diff}$ decreases with decreasing $Pr$. This is because the fluctuations within the BL become stronger as $Pr$ decreases, resulting in increased flux due to turbulence.

\begin{figure}
\centerline{\includegraphics[width=0.5\textwidth]{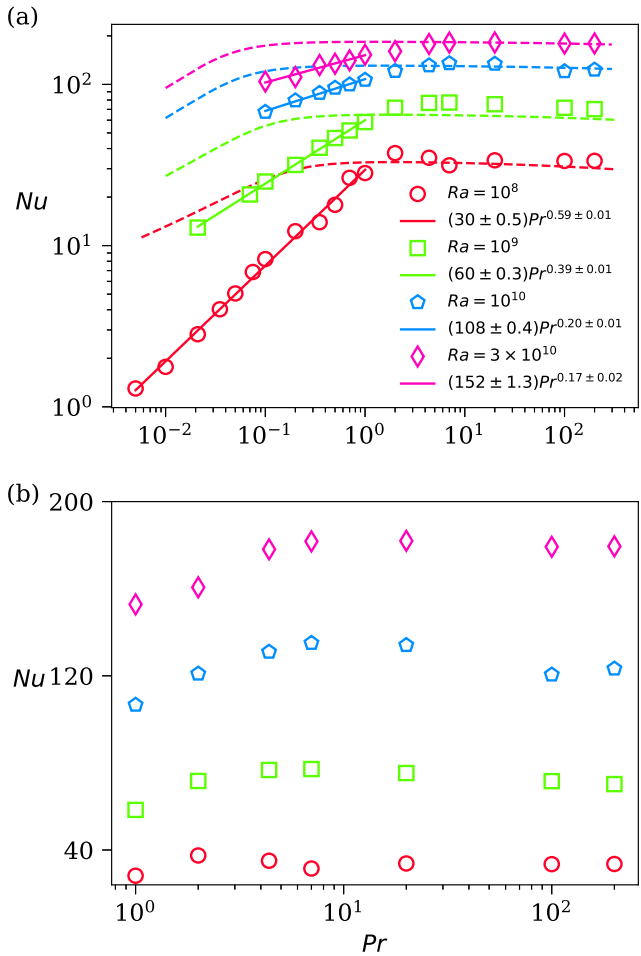}}
\caption{(a) The Nusselt number in the slender cell increases as $Pr^{\beta}$ for $Pr \leq 4$. The best fits for $Pr \leq 1$ yield power-law exponents $\beta$ that decrease with increasing $Ra$, approximately saturating to the value for $\Gamma \geq 1$ convection cells. Dashed curves represent $Nu$ computed using Grossmann-Lohse model~\cite{Stevens:JFM2013}. (b) $Nu$ for $Pr \geq 1$ shows that it first decreases slightly before leveling off for the two lower $Ra$ cases, but levels off without significant decrease at $Ra = 3 \times 10^{10}$.}
\label{fig:Nu_Pr}
\end{figure}

Although the convective component of the heat flux becomes stronger with decreasing $Pr$ within the BL region, the global heat transport decreases with decreasing $Pr$ for $Pr \leq 1$. The globally-averaged Nusselt number is computed, by integrating eq.~(\ref{eq:Nu_z}) in the vertical direction, as $Nu = 1 + \sqrt{Ra Pr} \, \langle u_z T \rangle_{V,t},$ where $\langle \cdot \rangle_{V,t}$ denotes averaging over the entire simulation domain and time. $Nu$ as a function of $Pr$ for all $Ra$ are plotted in fig.~\ref{fig:Nu_Pr}(a), which shows that $Nu$ increases with increasing $Pr$ up to $Pr \approx 2$ and becomes nearly independent of $Pr$ for larger $Pr$, which is consistent with earlier observations in $\Gamma \geq 0.5$ cells~\cite{Verzicco:JFM1999, Silano:JFM2010, Petschel:PRL2013, Poel:JFM2013, Pandey:Nature2018}. We find that $Nu$ increases as $Pr^{\beta}$ and the best fit  for $Pr \leq 1$ yields $\beta = 0.59, 0.39, 0.20, 0.17$ for $Ra = 10^8, 10^9, 10^{10}, 3 \times 10^{10}$, respectively. The exponents for $Ra = 10^8$ and $10^9$ are higher than those observed in larger-$\Gamma$ convection~\cite{Verzicco:JFM1999, Petschel:PRL2013, Scheel:PRF2017, Pandey:Nature2018}. As also observed in Iyer et al.~\cite{Iyer:PNAS2020}, we find that the magnitude of $Nu$ for $Pr \geq 1$ is very similar to that observed in larger-$\Gamma$ convection. Therefore, a larger $\beta$ implies that the global heat transport is reduced at lower $Pr$ for $Ra = 10^8$ and $10^9$. For instance, for $Pr = 0.021$ and $Ra = 10^8$, we observe $Nu = 2.8 \pm 0.05$, which is lower than $19.1 \pm 1.3$ for the corresponding parameters in a $\Gamma = 1$ cell~\cite{Scheel:PRF2017}. On the other hand, for $Pr$ = 4.38 and 20 at $Ra = 10^8$, Nusselt numbers in the slender cell are 35.1 and 33.8, respectively, which are close to 33.1 and 32.5 observed in the corresponding $\Gamma = 1$ cell by van der Poel et al.~\cite{Poel:JFM2013}. We also show Nusselt numbers computed using the Grossmann-Lohse (GL) model~\cite{Stevens:JFM2013} (dashed curves in fig.~\ref{fig:Nu_Pr}(a)), whose prediction of $Nu$ in cells of $\Gamma = 1$ is quite close to the present data for $\Gamma = 0.1$ when $Pr \geq 1$. Thus, the geometric confinement reduces turbulent heat transport in low-$Pr$ convection but not in high-$Pr$ cases. 

The reason for the reduced heat transport might be that the plumes become relatively thicker in the slender cell. For example, the thermal BL thickness for $Pr = 0.005, Ra = 10^8$, estimated by $0.5 H/Nu$, is nearly $0.4H$. Thus, the hot thermal plumes generated at the bottom plate cannot move horizontally due to severe confinement. They move upward and eventually collide with the cold sinking plumes, which inhibits the vertical movement of the plumes, eventually reducing heat transport. 

Figure~\ref{fig:Nu_Pr}(b) is an expanded view of the Nusselt number data for $Pr \geq 1$. For all Rayleigh numbers, the increase of $Nu$ at lower Prandtl numbers is followed by a slight decrease with increasing $Pr$~\cite{Xia:PRL2002} and a plateau. As mentioned earlier, the flow for $Ra = 10^8$ is time-independent for $Pr \geq 0.7$, and the Prandtl number at which $Nu$ levels off is about $7$. The same comments hold for $Ra = 10^9$. For $Ra = 10^{10}$, the approximate leveling off is observed at higher $Pr$, whereas for $Ra = 3 \times 10^{10}$, we observe the levelling off of the Nusselt number without an initial drop seen for lower $Ra$.  

The exponent $\beta$ for $Ra \geq 10^{10}$ saturates at about $0.19$. In a $\Gamma = 1$ cylindrical cell, Verzicco and Camussi~\cite{Verzicco:JFM1999} observed $\beta = 0.14$ for $Ra = 6 \times 10^5$; Scheel and Schumacher~\cite{Scheel:PRF2017} obtained $\beta = 0.11, 0.13, 0.15$ for $Ra = 3 \times 10^5, 10^6, 10^7$ respectively; and van der Poel et al.~\cite{Poel:JFM2013} observed  $\beta = 0.19$ for $Ra = 10^8$. In a horizontally-periodic cubic box, Petschel et al.~\cite{Petschel:PRL2013} observed $\beta = 0.19$ for $Ra = 5 \times 10^6$ and in a $\Gamma = 25$ box Pandey et al.~\cite{Pandey:Nature2018, Pandey:2021} observed $\beta \approx 0.19$ for $Ra = 10^5$ and $10^6$. Thus, one might fairly say that, for $Ra \geq 10^{10}$, $\beta$ in the slender cell approaches the value observed in $\Gamma \geq 1$ cells. Note that $Nu$ in $\Gamma > 1$ domains are slightly different from that in $\Gamma = 1$ cell~\cite{BailonCuba:JFM2010, Stevens:PRF2018}, while $Nu$ is found to be nearly independent of $\Gamma$ for $\Gamma > 4$~\cite{Stevens:PRF2018}. The precise value of the exponents $\alpha$ and $\beta$ may depend, in principle, on the range of $Ra$ and $Pr$ considered~\cite{Grossmann:JFM2000}. Our fits for $Nu$ in somewhat different ranges yield only slightly different numbers, so the numbers shown here are robust. 

\begin{figure}
\centerline{\includegraphics[width=0.47\textwidth]{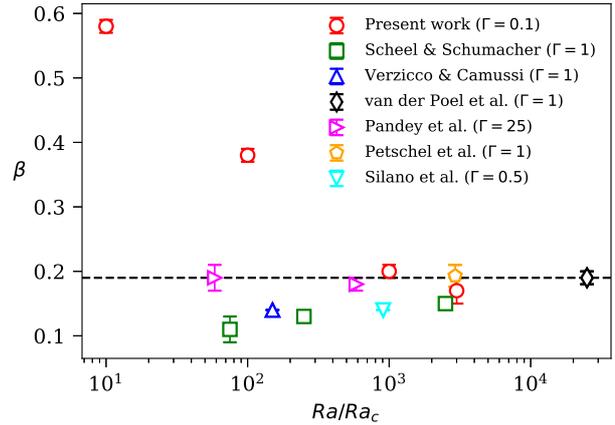}}
\caption{The power law exponent $\beta$ of the $Nu-Pr$ scaling as a function of $Ra$ normalized with the $\Gamma$-dependent critical Rayleigh number, $Ra_c$.  For $Ra/Ra_c \geq 10^3$, $\beta \approx 0.20$ is observed for convection in different aspect ratio cells. Data are taken from Refs.~\cite{Scheel:PRF2017, Verzicco:JFM1999, Poel:JFM2013, Pandey:Nature2018, Pandey:2021, Petschel:PRL2013, Silano:JFM2010}. The dashed line represents the exponent $\beta = 0.19$. The error bars are from best-fits.}
\label{fig:Nu_exponent}
\end{figure}

The critical Rayleigh number $Ra_c$ for the onset of convection increases with decreasing $\Gamma$~\cite{Yu:POF2017}. It is approximately $1.1 \times 10^7$ in the slender cell. Therefore, we observe no turbulent convection for $Pr \geq 0.7$ at $Ra = 10^8$. To compare the exponents $\beta$ in the slender cell with those in cells with $\Gamma \geq 0.5$, we plot them as a function of $Ra/Ra_c(\Gamma)$ in fig.~\ref{fig:Nu_exponent}, and note that $\beta$ in the slender cell approaches that in $\Gamma \geq 1$ cells for $Ra/Ra_c \approx 10^3$. Figure~\ref{fig:Nu_exponent} reveals that the asymptotic $\beta$ is attained in large $\Gamma$ cells at relatively low $Ra$ \cite{Pandey:Nature2018, Pandey:2021}, while it takes much higher $Ra$ for $\Gamma$ of the order unity (for which the approach is from below); for the slender cell, $\beta$ approaches the asymptotic value from above. From this perspective, one can claim that aspect ratios of the order 1 and below are characteristic of high aspect ratios only at very high $Ra$.

\begin{figure}
\centerline{\includegraphics[width=0.5\textwidth]{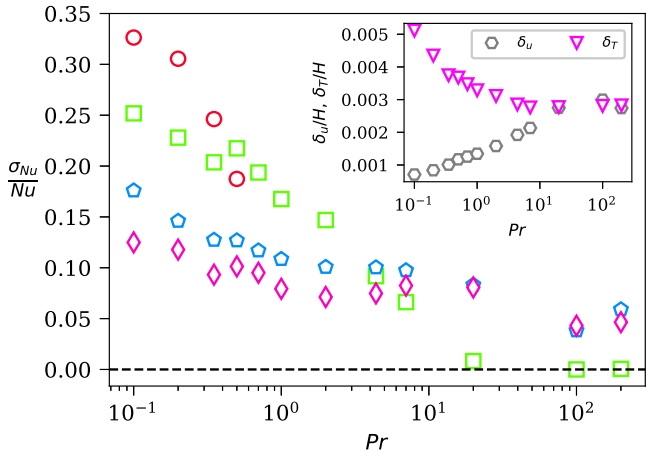}}
\caption{Relative standard deviation $\sigma_{Nu}/Nu$ of the heat flux at the plates increases with decreasing $Pr$. However, it decreases with increasing $Ra$ for low $Pr$. Symbols mean the same as in fig.~\ref{fig:Nu_Pr}. Inset shows the BL thicknesses as a function of $Pr$ for $Ra = 3 \times 10^{10}$.}
\label{fig:std_Nu}
\end{figure}

The heat transport at the plate is purely diffusive and we compute $Nu$ at the top and bottom plates at each time step in our simulations. We find good agreement between the time-averaged heat flux at the plates and the volume-averaged Nusselt number computed using the correlation $\langle u_z T \rangle$, which indicates that the statistics are sufficient for all simulations. Using the time series of $Nu$ at the plates, we compute the relative standard deviation, which is the ratio of the standard deviation to the mean of $Nu(t)$, and plot it as a function of $Pr$ in fig.~\ref{fig:std_Nu}. We observe that $\sigma_{Nu}/Nu$ increases with decreasing $Pr$, consistent with the increasing level of turbulence with increasing $Re$ or decreasing $Pr$. Further, the ratio at low $Pr$ decreases as $Ra$ increases due a rapid increase of the mean $Nu$ up to $Ra = 10^{10}$. We again note that the data for $Ra = 10^{10}$ and $3 \times 10^{10}$ are close.

We obtain estimates of the viscous BL thickness $\delta_u$ from the root-mean-square horizontal velocity $u_h(z) = \langle u_x^2 + u_y^2 \rangle^{1/2}_{A,t}$ using the slope method~\cite{Scheel:JFM2012}, and plot $\delta_u$ as well as $\delta_T \,\, (= 0.5H/Nu)$ for various $Pr$, fixing $Ra$ to be $3 \times 10^{10}$ (see the inset of fig.~\ref{fig:std_Nu}). The data show that $\delta_T$ decreases up to $Pr \approx 5$ and does not change appreciably thereafter. We also observe that $\delta_u$ increases with increasing $Pr$, barring the highest $Pr$ data point. Thus, the BL thicknesses for the highest $Pr$ for $Ra = 3 \times 10^{10}$ are about a hundredth of the horizontal dimension of the slender cell.

\section{Momentum transport}

\begin{figure}
\centerline{\includegraphics[width=0.5\textwidth]{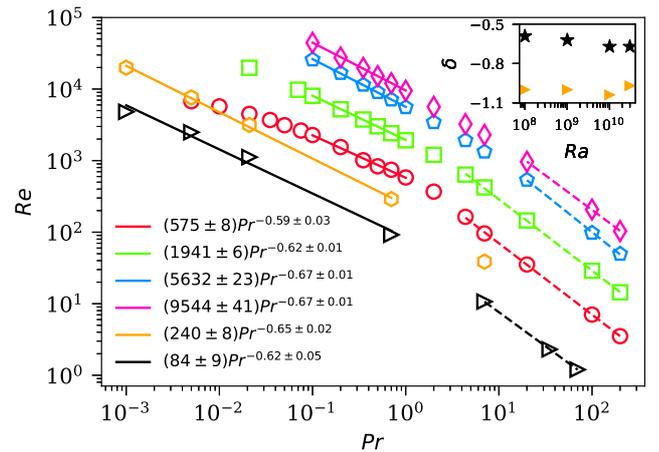}}
\caption{The Reynolds number as a function of $Pr$ for various $Ra$. Symbols mean the same as in fig.~\ref{fig:Nu_Pr}. $Re$ increases with decreasing $Pr$ and exhibits no clean power law of the type $Re \sim Pr^{-\delta}$. The best fit curves for $Pr$ between 0.1 and 1 (solid lines) for all the $Ra$ exhibit that $\delta$ nearly saturates for $Ra \geq 10^{10}$. The $\delta$ is larger for higher $Pr$ and the dashed lines are the best fits for the last few low-$Re$ points. The inset shows the power law exponent at low $Pr$ (black stars) and at high $Pr$ (orange triangles) as a function of $Ra$. Also shown for comparison are the data from simulations in a $\Gamma = 25$ box for $Ra = 10^5$ (black triangles) and $Ra = 10^6$ (orange hexagons)~\cite{Pandey:Nature2018, Pandey:2021}.}
\label{fig:Re_Pr}
\end{figure}

The turbulent momentum transport is quantified by the Reynolds number ($Re$), which can be defined in several ways depending on the definition of the length and velocity scales in the flow. In RBC, the height $H$ is usually chosen as the length scale and the recirculating motion of the large-scale as the velocity scale~\cite{Chilla:EPJE2012}. In DNS studies, the root-mean-square (rms) velocity is also often taken as the velocity scale. Here, we compute $Re$ using the rms velocity as 
\begin{equation}
\label{eq:Re}
Re = \sqrt{Ra/Pr} \, u_\mathrm{rms} = \sqrt{\langle u_i^2 \rangle_{V,t} \, Ra/Pr},
\end{equation}
and plot it as a function of $Pr$ in fig.~\ref{fig:Re_Pr}.
We find that $Re$ increases with decreasing $Pr$ as $Pr^{-\delta}$, i.e., the flow becomes increasingly turbulent as $Pr$ decreases. For $Ra = 10^8$, we find that $Re$ decreases slowly for $Pr$ between 0.005 and 0.035 and the best fit yields $\delta = 0.31 \pm 0.02$. However, $\delta$ increases for larger $Pr$ and becomes $0.57 \pm 0.01$ for $Pr$ between $0.035$ and 2. To compare $\delta$ for all $Ra$, we perform the best fit between $Pr =0.1$ and 1 and find $\delta = 0.59, 0.62, 0.67, 0.67$ for $Ra = 10^8, 10^9, 10^{10}, 3 \times 10^{10}$, respectively. Black stars in the inset of fig.~\ref{fig:Re_Pr} indicate the variation of $\delta$ with $Ra$. We also plot the Reynolds numbers from simulations in a $\Gamma = 25$ box at $Ra = 10^5$ and $10^6$~\cite{Pandey:Nature2018, Pandey:2021} in fig.~\ref{fig:Re_Pr} and find that the exponent $\delta$ for moderate Prandtl numbers lies in the aforementioned range in our slender cell. We also compute $Re$ based on the rms velocity in midplane at $z = H/2$ and find that it is a bit higher than $Re$ based on $u_\mathrm{rms}$ in the entire flow. However, the scaling exponent $\delta$ for all $Ra$ is very similar to that exhibited in fig.~\ref{fig:Re_Pr} (not shown here to avoid clutter).

In $\Gamma=1$ cylindrical cell~\cite{Verzicco:JFM1999, Scheel:PRF2017} and $\Gamma > 1$ rectangular box~\cite{Schmalzl:EPL2004, Pandey:Nature2018, Pandey:2021}, $\delta$ was found in the range $0.60-0.75$ for low Prandtl numbers. Grossmann-Lohse theory~\cite{Grossmann:JFM2000} also fits $\delta$ between $3/5$ and $2/3$ in a $\Gamma \approx 1$ domain. Thus, $\delta$ in our slender cell lies already in the observed range in $\Gamma \geq 1$ cell and saturates to $\delta \approx 0.67$. As in Iyer et al.~\cite{Iyer:PNAS2020}, we observe that the magnitude of $Re$ is lower compared to that in $\Gamma \approx 1$. This is because of larger drag from the sidewalls in the slender cell. 

Turbulent fluctuations become weaker with increasing $Pr$, and, for very high $Pr$, the buoyancy force is almost balanced by the viscous force due to weaker inertial force~\cite{Pandey:POF2016, Pandey:PRE2016}. Consequently, the Reynolds number decreases more rapidly with $Pr$ for higher $Pr$~\cite{Verzicco:JFM1999, Schmalzl:EPL2004, Pandey:Nature2018, Bhattacharya:POF2021}. For instance, Pandey et al.~\cite{Pandey:Nature2018} found $\delta \approx 0.96$ for $Pr \geq 7$ and $Ra = 10^5$ in RBC in a $\Gamma = 25$ box. We also observe a similar trend for the slender cell, and by fitting the last few low-$Re$ data points we find $\delta \approx -1.0 \pm 0.04$ for all the Rayleigh numbers. The best fits are exhibited as dashed lines in fig.~\ref{fig:Re_Pr} and the resulting exponents are plotted as orange triangles in the inset. Thus, the present exponents for higher $Pr$ too are consistent with those observed in $\Gamma \geq 1$ cells~\cite{Verzicco:JFM1999, Schmalzl:EPL2004, Pandey:Nature2018, Bhattacharya:POF2021}.

\section{Conclusions}

We studied the integral quantities for convection in a slender cylindrical cell of aspect ratio 0.1 by performing DNS for Prandtl numbers between 0.005 and 200 and Rayleigh numbers between $10^8$ and $3 \times 10^{10}$. We found that the Nusselt number in the slender cell is within 10\% to that in a $\Gamma = 1$ cell when $Pr$ is unity and larger. For lower Prandtl numbers, however, $Nu$ in the slender cell is lower than that for $\Gamma \sim 1$ cells. This lower heat transport leads to a steeper dependence on the Prandtl number. We found that the exponent $\beta$ in $Nu \sim Pr^{\beta}$ saturates approximately to 0.19 when $Ra$ is large, which is in the range of exponents observed in $\Gamma \geq 1$ convection flows. Our findings are that at high enough Rayleigh numbers and $Pr \geq 1$, heat transport is similar in slender and in higher aspect ratio cells within 10\%. Presumably this happens because the boundary layers become thin enough that the influence of the side boundary is negligible. Thus, what may matter for heat transport is less the aspect ratio but the ratio of the boundary layer thickness to the horizontal dimension of the convection cell.

As the convection in Solar and stellar interiors occur at very low $Pr$ and very high $Ra$~\cite{Schumacher:RMP2020}, a slender geometry would allow us to reach closer to these parameters with the presently available computational resources~\cite{Iyer:PNAS2020}. Numerical investigations of low-$Pr$ convection is very challenging~\cite{Schumacher:PNAS2015, Pandey:Nature2018}, which has prevented us to explore flows at very high $Ra$, but that is the direction in which we are proceeding at present. 

\acknowledgments
We thank J\"org Schumacher for many fruitful discussions and Richard Stevens for providing GL fits for $Ra = 10^{10}$ and $3 \times 10^{10}$ data. This research was carried out at the Center for Space Science at New York University's Abu Dhabi Campus using its High Performance Computing resources. The work was funded by the NYUAD Institute Grant G1502.

\end{document}